\documentclass[twocolumn,twoside,slac_two]{revtex4}
\usepackage{graphicx}
\usepackage{fancyhdr}
\begin{document}

\def \ups {{\mbox{$\Upsilon$}}}
\def \uone {{\mbox{$\Upsilon(1S)~$}}}
\def \utwo {{\mbox{$\Upsilon(2S)~$}}}
\def \uthree {{\mbox{$\Upsilon(3S)~$}}}
\def \ufour {{\mbox{$\Upsilon(4S)~$}}}
\def \jpsi {{\mbox{$J/\psi~$}}}
\def \chib {{\mbox{$\chi_b~$}}}
\def \chibpr {{\mbox{$\chi_b'~$}}}
\def \bmm {{\mbox{$B_{\mu\mu}~$}}}
\def \gee {{\mbox{$\Gamma_{ee}~$}}}
\def \geeuone {{\mbox{$\Gamma_{ee}(\Upsilon(1S))$}}}
\def \geeutwo {{\mbox{$\Gamma_{ee}(\Upsilon(2S))$}}}
\def \pipm {{\mbox{$\pi^+\pi^-~$}}}
\def \pipmz {{\mbox{$\pi^+\pi^-\pi^0~$}}}
\def \pizz {{\mbox{$\pi^0\pi^0~$}}}
\def \ee {{\mbox{$e^+e^-~$}}}
\def \mumu {{\mbox{$\mu^+\mu^-~$}}}
\def \mpipi {{\mbox{$M_{\pi\pi}~$}}}

\title{Upsilon Decays}


\author{Helmut Vogel}
\affiliation{Department of Physics, Carnegie Mellon University,
Pittsburgh, PA 15213}

\begin{abstract}
Bound upsilon states and their decays provide a unique laboratory
for testing QCD, LQCD, and quarkonium potential models. I review recent 
results, some of which are significant improvements in precision over
previous measurements and others are first-time observations. While
most of the results presented are from CLEO III there are also notable
contributions from BaBar and Belle who have just begun to exploit their
strong capabilities for studying bound state upsilons.

\end{abstract}

\maketitle

\thispagestyle{fancy}


\section{Introduction}
The physics to be extracted from properties and decays of bound state
upsilons is rich. The masses and widths of the various states and
cascade transitions between them test QCD, LQCD, and quarkonium
potential models. Inclusive production of charmonium in upsilon decay
tests QCD-inspired color-singlet and color-octet models. Radiative
decays to light hadrons test scaling between charmonium and bottomium;
they are also a source of glueballs should those exist. Searching for
LFV decays, $e.g.~\uone\to\tau\mu$, probes physics beyond the SM.

After a long period of near stagnation, the study of bound state
upsilons and their decays experienced an impressive renaissance,
starting in 2001-02 with the CLEO III experiment collecting dedicated
data samples at the \uone, \utwo, and \uthree resonances. The data
collected correspond to $21M$, $10M$, and $5M$ resonance decays,
respectively. In addition, BaBar and Belle have begun to analyze
their wealth of data also with a view towards bound upsilon states,
and have come out with exciting new results.

This report reviews highlights from the following subjects:
 \ (i) E1 photon transitions, \ (ii) ``unusual'' hadronic transitions,
 \ (iii) \uone decay to charmonium, \ (iv) radiative decays of the \uone,
 \ (v) precision measurements of \bmm and \gee , and (vi) dipionic 
cascade transitions-- an old topic on which CLEO III as well as BaBar
and Belle have just shed new light.

\section{E1 photon transitions}

Photon transitions between bound upsilon states probe quarkonium
potential models. In particular, measurement of transition energies
and branching fractions probe the spin dependence and the magnitude of
relativistic corrections. CLEO III has recently performed a definitive
study\cite{cleo_e1_1} of the E1 transitions, $\utwo\to\gamma\chib$,
$\uthree\to\gamma\chib$, and $\uthree\to\gamma\chibpr$. As one example
of the sensitivity achieved, the branching fraction for the decay
$\uthree\to\gamma\chib(^1P_0)$ was measured to be
$(0.30\pm0.04\pm0.10)$\%. This allowed to discriminate between various
theoretical models the predictions of which ranged between 0.006 and
0.74\%. Another highlight has been the discovery of the $1^3D_J$
state\cite{cleo_e1_2}, the first new upsilon state in 20 years and the
first long-lived $L=2$ meson below open-flavor threshold. The mass
difference between this $1D$ state and the \uone turned out to be one
of ten ``golden'' experimental input quantities against which to test
new, unquenched LQCD calculations\cite{lattice}. This test has been
spectacularly successful.

\section{``Unusual'' hadronic transitions}

Hadronic cascade transitions between upsilon states test models of
soft gluon emission with subsequent hadronization into light hadrons
\cite{tmyan}.  From the late 1970s until recently, the
only observed such hadronic cascades had been dipion transitions
between triplet S states. Now, results on two different types of
hadronic cascades have been reported.

\subsection{$\chibpr \to \omega \uone$}

This is the first observation of a non-pionic cascade transition
between $b\bar{b}$ states, made by CLEO III via the channel,
$\uthree\to\gamma\chibpr, ~ \chibpr\to\omega\uone$, with
$\omega\to\pipmz$ and $\uone\to l^+l^-$. The final state consists
of $\gamma + \pipmz + l^+l^-$.  The invariant \pipmz mass
distribution shows a clean $\omega$ signal. The energy distribution of
the photon associated with the $\omega$ shows two peaks corresponding,
respectively, to the $J=2$ and $J=1$ levels of \chibpr (the cascade
via the $J=0$ level is kinematically forbidden). CLEO III measures the
branching fractions for $\chi_{bJ}'\to\omega\uone$ to be 1.6\% and
1.1\% for $J=1$ and $J=2$, respectively\cite{cleo_omega}. Considering
the tight available phase space these branching fractions are very
large, in agreement with an early prediction by
Gottfried\cite{gottfried}. A discussion of the spin dependence is
given by Voloshin\cite{voloshin}.

\subsection{$\chibpr \to \pi\pi \chib$}

This is the first observation of a dipion cascade between non-S states,
made by CLEO III via  
the cascade, $\uthree\to\gamma\chibpr, ~ \chibpr\to\pi\pi\chib,
~ \chib\to\gamma\uone$, with $\uone\to l^+l^-$.
The final state consists of $2\gamma + 2\pi + l^+l^-$.
The experimental challenge lies in the fact that the pions are very soft
and that the main background process, $\uthree\to\pi\pi\utwo$ with
$\utwo\to\gamma\chib$ where $\chib\to\gamma\uone$ as before, has
identical final state particles with very similar kinematics. CLEO III
applied two analysis methods, one in which both pions are detected
(low statistics/low background) and one in which only one pion is
detected and the other one is inferred from kinematic constraints
(higher statistics/higher background). They observe the cascade transition 
as a $6\sigma$ effect\cite{cleo_chibpr} and obtain the partial width, \ 
$\Gamma_{\pi\pi}(\chibpr) = (0.83\pm0.22\pm0.08\pm0.19)~keV$, consistent
with the prediction of \ $\Gamma_{\pi\pi}(\chibpr)\approx 0.4~keV$ by
Kuang \& Yan \cite{kuangyan}.

\section{Inclusive decay to charmonium, $\uone\to\jpsi~X$}

This decay tests models of charmonium production in gluon-rich
environments, in particular the color-octet\cite{coloroctet} and
color-singlet\cite{colorsinglet} models, both of which predict a
branching fraction and the shape of the inclusive \jpsi momentum
distribution. CLEO III studied this process using
their data sample of $21M$ \uone which is a nearly twentyfold increase
in statistics over the only earlier measurement, by CLEO II, of more
than a decade ago.  CLEO III select events with inclusive high-momentum
lepton pairs. They observe clean \jpsi signals in both the \ee and
the \mumu channel \cite{cleo_jpsi}. The measured branching fraction,
\cal{B}$(\uone\to\jpsi X) = (6.4\pm0.4\pm0.6)\times10^{-4}$, favors
the color-octet model. The observed \jpsi momentum spectrum, however,
is significantly softer than predicted by the color-octet model and
also softer -- although closer to -- that predicted by the
color-singlet model, indicating final state interactions not included
in either model. Inclusive production of $\psi(2S)$ and $\chi_c$ in
\uone decay is also observed.

\section{Radiative decays of \uone}

\subsection{$\uone \to \gamma h^+h^-,~~(h=\pi,K,p)$}

In this type of decay the light hadron pair is produced in a gluon-rich
environment. It is an ideal source of glueballs should those exist. 
Scaling from the \jpsi where these decays have been studied extensively,
to the \uone the rates are expected to be suppressed by a factor of
$((q_b m_c)/(q_c m_b))^2\approx 1/40$ and the branching fractions
by a factor of $\approx 1/25$. CLEO III has studied these decays,
as well as the $\gamma\pizz$ final state which is free of $\gamma\rho$
continuum background. They fit the observed struture in the $hh$ 
invariant mass distribution with relativistic, spin-dependent Breit-Wigner
curves. The main results \cite{cleo_ghh} are the following. (i) The
$f_2(1270)$ is confirmed in the $\pi\pi$ channel, and from a fit within 
the helicity formalism the spin assignment, $J=2$, is established.
(ii) The $f_2'(1525)$ is observed in the $KK$ channel, also as $J=2$.
(iii) The rates are consistent with expectations based on scaling from
the \jpsi. (iv) Tensor mesons dominate the observed structure. (v)
No signal is found for $f_J(2200)$ in any of the final states, leading
to upper limits on the product branching fractions of the order 
of \ $6-11\times 10^{-7}$.

\subsection{Search for $\uone \to \gamma \eta(')$}

These processes are theoretically simple in that there are no hadronic
final-state interactions. They have been studied extensively in \jpsi
radiative decay where good agreement with theory was found. Studying
these decays in the upsilon system tests models of scaling such as VDM,
NRQCD, and mixing with $\eta_b$. The previous upper limit (from CLEO II)
on the branching fraction has stood at $\approx 2\times 10^{-5}$.
CLEO III chooses the 3 main decay modes for the $\eta$ and 4 modes for
the $\eta'$ ($3\times (\eta\pipm)$ and $\gamma\rho$). No signal is found
in any of the channels in $21M$ \uone events, leading to upper limits,
 \ \cal{B}$(\uone\to\gamma\eta) < 9.3\times 10^{-7}$ \ and
 \ \cal{B}$(\uone\to\gamma\eta') < 17.7\times 10^{-7}$.
This strongly disfavors mixing with $\eta_b$. It is still consistent
with VDM and, marginally, with NRQCD. These results are preliminary.

\section{Precision measurement of \bmm and \gee} 

The muonic branching fraction and the di-electron width of a quarkonium
resonance are fundamental quantities in themselves. Beyond that, \bmm
frequently enters into derivation of other branching fractions while
\gee is important for comparisons with LQCD. CLEO III has re-measured
\bmm and \gee for the bound state upsilon resonances with high precision.

\subsection{\bmm of \uone, \utwo, and \uthree}

Here, the main experimental challenges are precision luminosity
measurement, subtraction of the copiously produced continuum muon
pairs, and suppression of cosmic ray background. CLEO III
reports\cite{cleo_bmm} the new measurements,
$\bmm(\uone)=(2.49\pm0.02\pm0.07)$\%,
$\bmm(\utwo)=(2.03\pm0.03\pm0.08)$\%, and
$\bmm(\uthree)=(2.39\pm0.07\pm0.10)$\%.  This corresponds to a
relative precision of $2-3$\% in each case, far exceeding that 
any previous measurement. Note also that while the value of \bmm
for \uone is consistent with previous measurements the results for
\utwo and \uthree are significantly higher than the previous PDG
values. This implies a downward shift in the values for the total
widths of \utwo and \uthree.

\subsection{\gee of \uone, \utwo, and \uthree}

\begin{figure}[h]
\centering
\includegraphics[width=80mm]{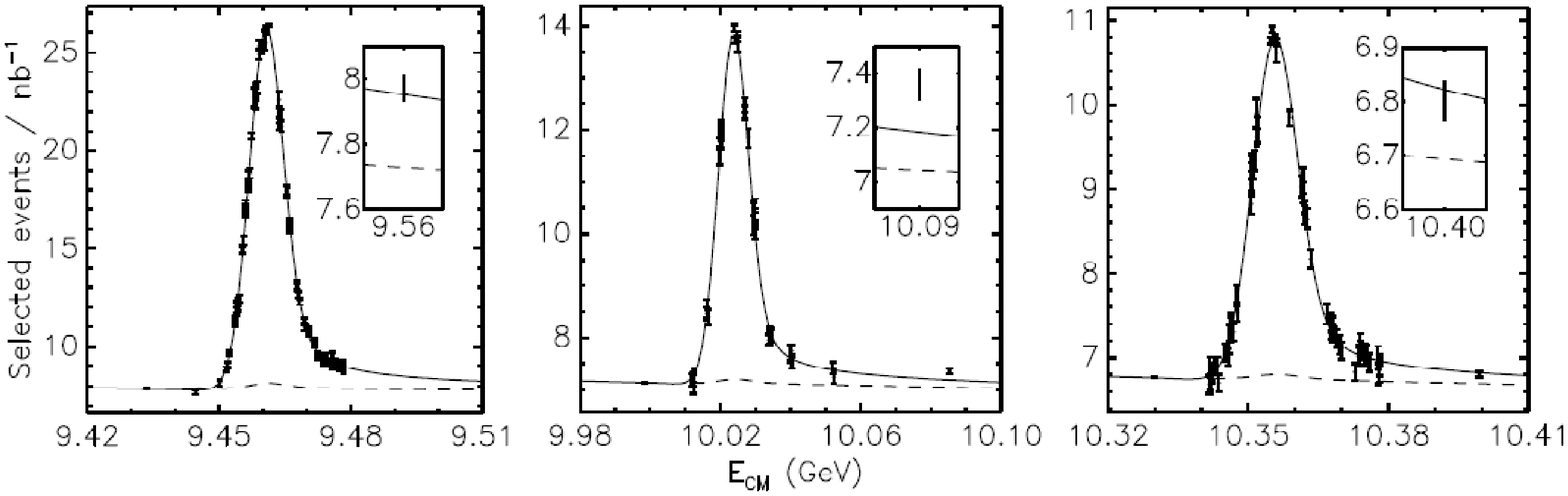}
\caption{CLEO III scans of the \uone, \utwo, and \uthree resonances.} 
\label{gee_scans}
\end{figure}

\begin{figure}[h]
\centering
\includegraphics[width=80mm]{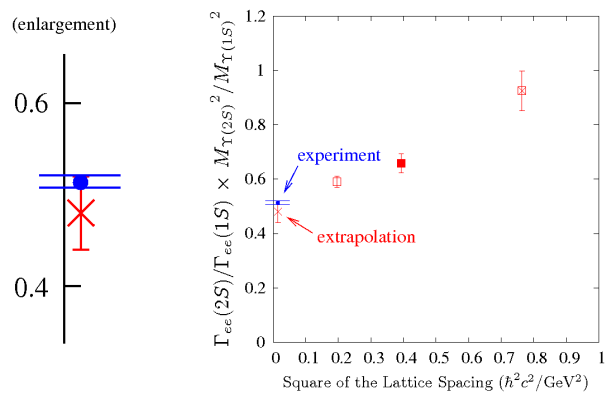}
\caption{Comparison of LQCD calculations\cite{lqcd_gee} with experimental 
results\cite{cleo_gee} on \geeutwo/\geeuone~ from CLEO III. Shown
on the left is a close-up of the region near zero lattice spacing.} 
\label{gee_lqcd}
\end{figure}

The di-electron width, \gee, of a bound upsilon state is proportional
to the integral of the total cross section when scanning across the
resonance in \ee collisions. (The width cannot be ``read off''
directly because it is only of the order of $keV$ whereas the observed
resonance shape is dominated by the beam energy spread of typically
several $MeV$.) The experimental challenge lies in tracking both the
height of the resonance (luminosity measurement, efficiency,
backgrounds) and its width (shift in beam energy). Effects
contributing to the observed lineshape include continuum production of
hadrons, two-photon fusion, cosmic-ray backgrounds, beam-gas
background, and the radiative tails of lower mass resonances. CLEO III
performed and analyzed repeated scans across each of \uone, \utwo, and
\uthree (11, 6, and 7, respectively, see figure~\ref{gee_scans}). 
The final results are\cite{cleo_gee}, \
$\gee(\uone)=(1.354\pm0.005\pm0.020)~keV$, \
$\gee(\utwo)=(0.619\pm0.007\pm0.009)~keV$, and \
$\gee(\uthree)=(0.446\pm0.004\pm0.007)~keV$.  The results are
consistent with and significantly more precise than any previous
measurements. This allows a stringent test of LQCD
where the quantities most reliably calculable are the ratios of
di-electron widths.  Figure~\ref{gee_lqcd} shows one such comparison
between experiment and extrapolated LQCD calculations\cite{lqcd_gee}.
The agreement is good, and also presents a challenge to the LQCD
community to further reduce the theoretical uncertainty.

\section{Dipion cascades between $\Upsilon(nS)$ states}

\begin{figure*}[t]
\centering
\includegraphics[width=150mm]{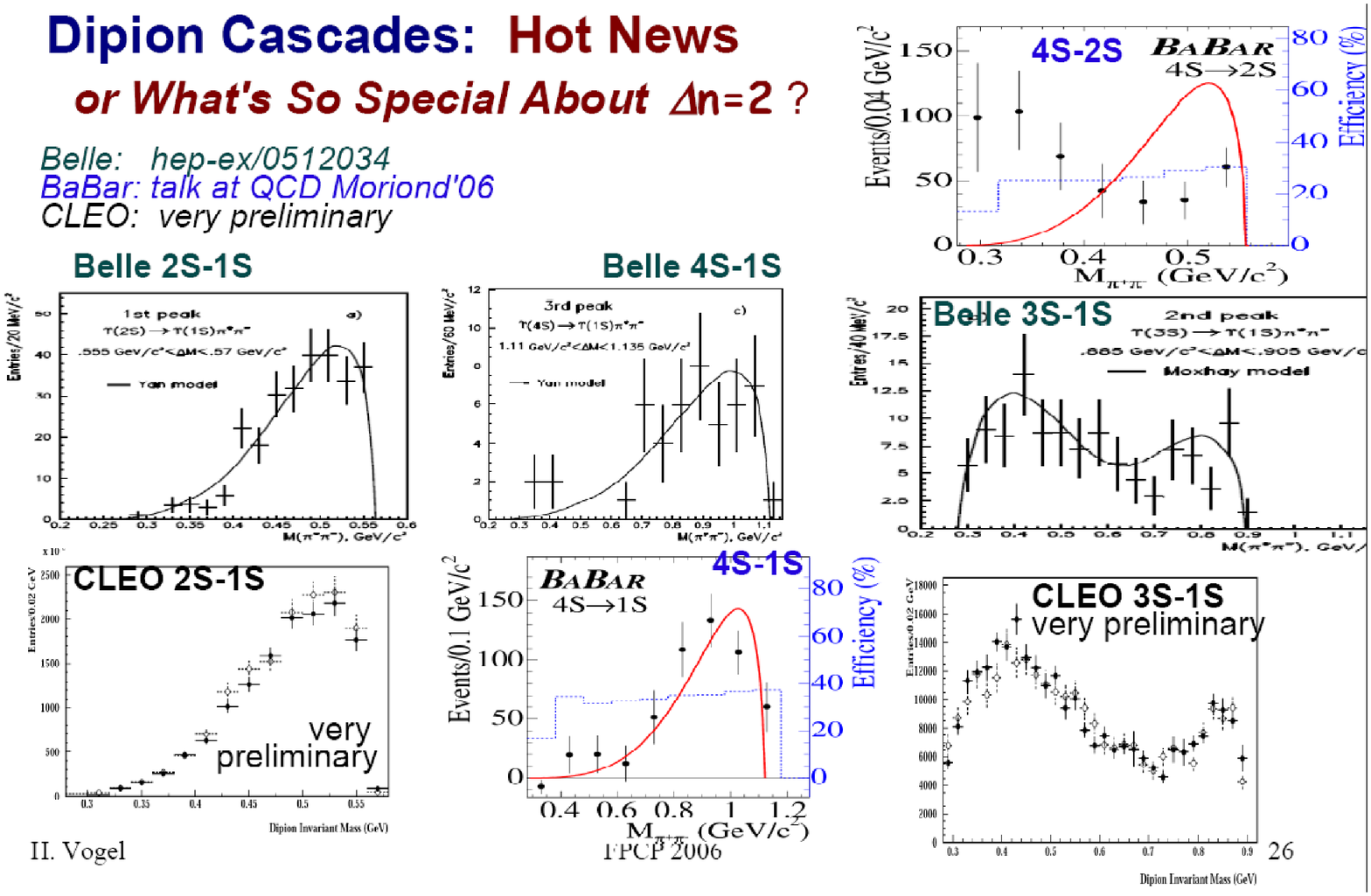}
\caption{Compilation of various recently measured \mpipi distributions 
of dipion cascade transitions between triplet S states in the
upsilon system. The data are from Babar, Belle, and CLEO III.} 
\label{all_mpipi}
\end{figure*}

In this paragraph, $nS\to mS$ denotes dipion cascade transitions
between triplet $S$ states in bottomium, $\ups(nS)\to\pi\pi\ups(mS)$.
Measurements of these cascades date back to the very early days of the
LENA, Argus, Crystal Ball, CUSB, and CLEO I experiments.  Quantities
of interest are the branching fraction and, in particular, the
distribution of the dipion invariant mass, \mpipi. It has long been
established that in the transition $2S\to1S$, \mpipi peaks towards the
upper end of the kinematic range just as in the corresponding
transition in charmonium (conforming to the Yan model\cite{tmyan}).
By contrast, in $3S\to1S$ \ \mpipi exhibits double peak structure
suggestive of final-state interactions, intermediate resonance
formation, and/or coupled-channel effects as in the Moxhay
model\cite{moxhay}). After nearly two decades of stagnation this
subject has recently found renewed interest with experimental input
from CLEO III as well as from BaBar and Belle.  CLEO III is in the
process of finalizing a high-statistics study of the $2S\to1S$,
$3S\to1S$, and $3S\to2S$ cascades. Babar just
reported\cite{babar_mpipi} results on $4S\to1S$ and $4S\to2S$
cascades. Belle also reported\cite{belle_mpipi} results on $4S\to1S$
and, from their ISR data sample results on $3S\to1S$ and
$2S\to1S$. The picture emerging from these results is intriguing, as
shown in figure~\ref{all_mpipi}.  Evidently,
\ $\Delta n = 2$ \ transitions exhibit double peak structure whereas \
$\Delta n \neq 2$ \ transitions do not.  At present there does not
appear to be a theoretical explanation for such pattern.

\section{Summary}
The study of properties and decays of bound upsilon states has
recently gained renewed interest. On the experimental side, results
from CLEO III have been either first time observations or
re-measurements which set new standards of precision. In addition,
BaBar and Belle have begun to exploit the wealth of bound state
upsilon data produced as ``byproduct'' of running on the \ufour, and
have already weighed in with exciting new results on dipion cascade
transitions.  Furthermore, Belle has done dedicated running at the
\uthree and may do some more in the near future. The results provide
key inputs to theory including quarkonium potential models, QCD and,
in particular, to LQCD where calculations are approaching the few
percent level in precision.

\bigskip % extra skip inserted
\begin{acknowledgments}
I thank my colleagues in the CLEO collaboration for their help with
preparing this talk. I am also grateful to Riccardo Faccini of BaBar
and Tom Browder of Belle for valuable input from their respective
experiments, and to George Hou for an informative discussion on
color-singlet $vs.$~color-octet models.  This work was supported by
the National Science Foundation and the U.S.~Department of Energy.
\end{acknowledgments}

\bigskip % extra skip inserted

\end{document}